%% file: arxiv.tex
\documentclass[10pt,conference]{IEEEtran}

\usepackage{cite}
\usepackage{amsmath,amssymb,amsfonts}
\usepackage{algorithmic}
\usepackage{graphicx}
\usepackage{textcomp}
\usepackage{xcolor}
\usepackage{balance}
\usepackage[all]{nowidow}
\usepackage[utf8]{inputenc}
\usepackage{newunicodechar}
\usepackage{acronym}
\usepackage{booktabs}

\usepackage[hidelinks]{hyperref}
\usepackage[capitalise]{cleveref}

\DeclareRobustCommand{\okina}{%
  \raisebox{\dimexpr\fontcharht\font`A-\height}{%
    \scalebox{0.8}{`}%
  }%
}
\newunicodechar{ʻ}{\okina}

\acrodef{YCSB}{\emph{Yahoo! Cloud Serving Benchmark}}
\acrodef{SBC}{single-board computer}
\acrodef{AWS}{Amazon Web Services}
\acrodef{SUT}{system under test}
\acrodef{LSTM}{long short-term memory}
\acrodef{ML}{machine learning}
\acrodef{QoS}{quality-of-service}
\acrodef{SLO}{service level objective}
\acrodef{IoT}{Internet of Things}
\acrodef{FaaS}{Function-as-a-Service}
\acrodef{usgs}[USGS]{\emph{U.S.~Geological Survey}}
\acrodef{hvo}[USGS HVO]{\emph{USGS Hawaiian Volcano Observatory}}

\begin{document}

\title{Towards a Benchmark for Fog Data Processing}

\author{\IEEEauthorblockN{Tobias Pfandzelter, David Bermbach}
    \IEEEauthorblockA{\textit{Technische Universit\"at Berlin \& Einstein Center Digital Future}\\
        \textit{Mobile Cloud Computing Research Group} \\
        \{tp,db\}@mcc.tu-berlin.de}
}

\maketitle

\begin{abstract}
    Fog data processing systems provide key abstractions to manage data and event processing in the geo-distributed and heterogeneous fog environment.
    The lack of standardized benchmarks for such systems, however, hinders their development and deployment, as different approaches cannot be compared quantitatively.
    Existing cloud data benchmarks are inadequate for fog computing, as their focus on workload specification ignores the tight integration of application and infrastructure inherent in fog computing.

    In this paper, we outline an approach to a fog-native data processing benchmark that combines workload specifications with infrastructure specifications.
    This holistic approach allows researchers and engineers to quantify how a software approach performs for a given workload on given infrastructure.
    Further, by basing our benchmark in a realistic IoT sensor network scenario, we can combine paradigms such as low-latency event processing, machine learning inference, and offline data analytics, and analyze the performance impact of their interplay in a fog data processing system.
\end{abstract}


\input{sections/1_introduction.tex}
\input{sections/2_background.tex}
\input{sections/3_requirements.tex}
\input{sections/4_scenario.tex}
\input{sections/5_benchmark.tex}
\input{sections/6_conclusion.tex}

\section*{Acknowledgements}

Funded by the Deutsche Forschungsgemeinschaft (DFG, German Research Foundation) -- 415899119.

\balance

\bibliographystyle{IEEEtran}
\bibliography{bibliography.bib}

\end{document}

%% file: sections/1_introduction.tex
\section{Introduction}
\label{sec:introduction}

With the advent of emerging application domains such as the \ac{IoT}, autonomous vehicles, and extended realities and metaverses, the fog computing paradigm has received increasing attention in research~\cite{bonomi2012fog,paper_bermbach_fog_vision}.
Fog computing extends the infinite resources of the cloud towards the edge of the network, with geo-distributed compute nodes closer to clients and devices~\cite{bonomi2012fog}.
This can reduce network strain on the backhaul network, with local data processing and aggregation, reduce latency for edge services given the physical proximity of resources, and increase privacy as centralized cloud data pools are no longer necessary~\cite{bonomi2012fog,paper_bermbach_fog_vision,paper_pfandzelter_functions_vs_streams,paper_pfandzelter_tinyfaas,paper_bermbach_auctions4function_placement,paper_bermbach_auctionwhisk}.

Managing compute and storage resources throughout the fog is no trivial task, as, e.g., data replication and \ac{QoS} guarantees must be managed~\cite{paper_hasenburg_towards_fbase,techreport_hasenburg_fbase,pfandzelter2023fred}.
Researchers have thus proposed abstractions such as platforms and middlewares for fog application development~\cite{pfandzelter2023fred,paper_pfandzelter_LEO_serverless,paper_pfandzelter_tinyfaas,8603163,rausch2019towards,rausch2019system,paper_hasenburg_geobroker,paper_hasenburg_geobroker_software,paper_hasenburg_disgb,paper_pfandzelter_coordination_middleware}.

Currently, there is no way to quantitatively evaluate how these approaches perform, especially \emph{across compute paradigms}.
We are lacking benchmarking approaches that (i)~can reflect the geo-distributed, multi-paradigm nature of fog data processing applications, (ii)~reflect the interdependence of infrastructure and applications in fog computing, and (iii)~allow us to compare different operational approaches to application management and deployment in the fog.

In this paper, we outline a benchmark architecture for quantitatively evaluating the performance of fog data processing abstractions.
We present an overview of existing work in geo-distributed data processing systems and their quantitative evaluation in \cref{sec:background}.
We further analyze the design objectives for our benchmark in \cref{sec:requirements}.
In \cref{sec:scenario}, we then present a fog application scenario around which our benchmark is built.
We describe benchmark workload, parameters, infrastructure, and metrics in \cref{sec:benchmark} and present initial implementation artifacts in \cref{sec:implementation}.

%% file: sections/2_background.tex
\section{Background \& Related Work}
\label{sec:background}

Before describing details of the benchmark we propose in this paper, we give an overview of fog data processing (\cref{sec:background:fogdata}) and related work on performance evaluation of geo-distributed data-intensive systems (\cref{sec:background:related}).

\subsection{Fog Data Processing}
\label{sec:background:fogdata}

Fog computing is an abstraction around the combined compute and storage resources at the edge, in the core network, and in the cloud~\cite{bonomi2012fog,paper_bermbach_fog_vision}.
Fog computing is a promising paradigm for emerging domains such as the \ac{IoT}:
By extending cloud resources close to the edge of the network, where data is generated, fog applications can benefit from low-latency, high-bandwidth, privacy-preserving data processing infrastructure~\cite{paper_pfandzelter_functions_vs_streams}.
For example, \ac{IoT} sensor data can be processed by compute resources at a local radio gateway and sent to actuators at the edge without incurring a significant network delay~\cite{paper_pfandzelter_tinyfaas}.
Similarly, data from multiple sensors can be aggregated and filtered at the edge in order to limit the network strain of sending all data to the cloud.
Personally identifiable data can be anonymized using on-device resources before it is further processed by a third party~\cite{paper_pallas_fog4privacy}.

The downside of geo-distributed, heterogeneous fog infrastructure is the complexity of its management for applications, i.e., building approaches for distributing data, deploying services, and managing fault-tolerance~\cite{paper_pfandzelter_LEO_serverless}.
Researchers have proposed abstractions in the form of fog data processing platforms that manage this complexity for applications.
For example, \emph{NebulaStream}~\cite{zeuch2020nebulastream,zeuch2020nebulastream2} is an end-to-end \ac{IoT} data management system.
A key novelty of NebulaStream is its ability to autonomously deploy different data processing approaches such as stream and complex event processing in a geo-distributed environment.
\emph{SoFA}~\cite{maleki2019sofa} uses Apache Spark to combine all available fog resources for stream processing operator deployment.
\emph{Lotus}~\cite{wang2023lotus} combines a fog publish/subscribe system with event and message processing based on the \ac{FaaS} paradigm~\cite{paper_pfandzelter_tinyfaas}.
In previous work, we have introduced the \emph{FReD}~\cite{pfandzelter2023fred,paper_hasenburg_towards_fbase,techreport_hasenburg_fbase} fog data management system that distributes and stores geo-distributed data with application-controlled replica placement, integrating distributed compute services.
All these approaches \textit{functionally} deliver the same data processing logic yet approach it in entirely different ways, which are hard to compare using traditional benchmarks.
The question of how they differ in QoS remains open.

\subsection{Related Work}
\label{sec:background:related}

Numerous benchmarks in the area of fog computing have been proposed.
In their survey on edge performance benchmarking, Varghese et al.~\cite{varghese2021survey} find that these benchmarks focus mostly on infrastructure components for the edge, e.g., comparing different CPUs or single-board computers for their suitability for edge processing.
For example, Morabito~\cite{morabito2017virtualization} evaluates the performance of \ac{IoT} edge devices along metrics such as power consumptions and task execution time.

In contrast, software platforms are considered only to a limited extent.
Confais et al.~\cite{confais2016performance} deploy a fog-like multi-site cluster of machines with \emph{Cassandra}, \emph{Rados}, and \emph{IPFS} database systems and run the cloud benchmark \ac{YCSB}~\cite{cooper2010benchmarking} against them.
The results show that all database systems underperform, as neither the workload nor the systems are built for a fog environment.

Das et al.~\cite{das2018edgebench} propose \emph{EdgeBench}, a benchmarking suite for serverless edge computing.
The authors use EdgeBench to evaluate commercial edge computing platforms from \acs{AWS} and Microsoft Azure.
Similar to other benchmarking studies by Gorlatova et al.~\cite{gorlatova2018characterizing}, Silva et al.~\cite{silva2019towards}, McChesney et al.~\cite{mcchesney2019defog}, George et al.~\cite{george2020openrtist}, and Verma et al.~\cite{verma2021edge}, EdgeBench focuses on a pipeline from edge device to cloud with only a single processing task moved between the fog layers.
This simplified view leaves out broader performance implications of device heterogeneity at the edge, multiple layers of intermediate resources, and interplay between fog services.

A more complex benchmark requires realistic applications and datasets, execution models for distributed workload testing, virtualization and emulation approaches for distributed fog environments, and meaningful metrics.
Tocz{\'e} et al.~\cite{tocze2019towards,tocze2022edge} propose a methodology for characterizing edge workloads and compile workload traces.
Kolosov et al.~\cite{kolosov2020benchmarking} compose realistic edge datasets using workloads from other areas.
Aslanpour et al.~\cite{aslanpour2020performance} review performance metrics for fog computing benchmarks.
Raith et al.~\cite{9946322} introduce \emph{Galileo}, an end-to-end experiment framework for distributed load testing in edge computing environments.
With Galileo, researchers can define and execute experiments, operationalize them on cluster testbeds, and instrument and collect monitoring data.
Mu{\~n}oz et al.~\cite{munoz2022cleave} propose \emph{CLEAVE}, a framework for performing experiments on Networked Control Systems at the edge.

Hasenburg et al.~present \emph{MockFog}~\cite{paper_hasenburg_mockfog,paper_hasenburg_mockfog2}, which provides an approach to build virtual testbeds of fog infrastructure in the cloud.
MockFog emulates fog nodes as cloud virtual machine instances and emulates network connections between those instances, including network delay, bandwidth constraints, and packet loss.
MockFog also provides a framework for experiment planning and control, simplifying testing and evaluation distributed fog applications.
However, MockFog makes no assumptions about the kind of testing that is performed and does not provide a complete fog benchmark.
Similarly, \emph{Celestial}~\cite{paper_pfandzelter2022celestial,techreport_pfandzelter_celestial_extended} emulates infrastructure for experimenting with LEO edge applications but does not provide a benchmark.

%% file: sections/3_requirements.tex
\section{Design Objectives}
\label{sec:requirements}

The design of a systems benchmark must meet a number of requirements, namely \emph{relevance}, \emph{repeatability} and \emph{reproducibility}, \emph{fairness}, and \emph{portability}~\cite{huppler2009art,book_cloud_service_benchmarking}.
For the fog data management and processing systems we plan to target with our benchmark, relevance means that the workload and parameter sets for the systems are representative of ``real'' fog workloads.
As we have yet to see fog infrastructure and software widely deployed in industry, we adopt the notions of these realistic fog workloads from research.

Notably, we observe three key aspects of fog workloads that must be mirrored in the benchmark design:
First, fog applications and middlewares are inherently tied to the infrastructure they are deployed on, given the importance of geo-distributed and heterogeneous software deployment in the fog~\cite{paper_pfandzelter_zero2fog}.
Second, fog applications are composed of interconnected services and functionalities that interact with data in various ways~\cite{paper_pfandzelter_functions_vs_streams}.
Third, how these different services are implemented depends heavily on the abstractions that fog middlewares provide, e.g., both stream processing engines and serverless platforms can be used to implement an online monitoring application~\cite{paper_pfandzelter_functions_vs_streams,paper_pfandzelter2022streamingfunctions}.
Based on these three aspects, we derive three design objectives that we aim to meet with our benchmark specification:

\subsubsection*{\textbf{Objective~1} Geo-Distributed Multi-Paradigm Workloads}

Fog applications comprise different services that interact with data in different ways, e.g., at the edge, \ac{IoT} sensors make local decisions on the data they generate with a focus on low latency data processing of individual data items.
In intermediate servers within the core network, data streams of multiple devices may be correlated in an online manner.
In the cloud, large volumes of offline and historical data are analyzed with a focus on throughput and efficiency.
Crucially, these services are interconnected, with edge on-device processing latency impacting aggregation delay in intermediary nodes, which in turn impacts data staleness and availability in the cloud.
We thus need our benchmark to reflect this interdependency of different services, with different data processing paradigms.
This also requires managing distributed clients and workloads of different kinds and combining and correlating their results.

\subsubsection*{\textbf{Objective~2} Combine Workload and Infrastructure Specification}

Cloud service benchmarks such as \ac{YCSB}~\cite{cooper2010benchmarking} treat the \ac{SUT} as an atomic entity, e.g., an API that is called.
While the infrastructure for a benchmark is fixed, e.g., a type of cloud virtual machine instance, it is not part of the benchmark specification.
A fog data processing benchmark, however, must define this infrastructure as part of the workload, in the sense that the \ac{SUT} must be able to deal with this infrastructure, i.e., run on what is available.
This creates two relevant specifications in our benchmark: the workload that is sent against the \ac{SUT} (\emph{application} perspective), and the infrastructure specification of the fog network, available compute resources, and network topologies (\emph{infrastructure} perspective).
Benchmark implementations must thus implement both the client workloads and the fog infrastructure, and quality metrics for both perspectives can be developed.

\subsubsection*{\textbf{Objective~3} Cross-Paradigm Portability}

Different paradigms for organizing data processing applications have been proposed to tackle the challenges of fog computing, e.g., container orchestration and Function-as-a-Service (FaaS)~\cite{paper_pfandzelter_LEO_serverless,paper_pfandzelter_zero2fog,paper_pfandzelter_functions_vs_streams}.
In an effort to maintain a fair comparison, application logic specified in the benchmark should be independent of processing paradigms.
The implementation of these services for a specific benchmark implementation will also impact how the \ac{SUT} performs, as compute resources occupied by the service will not be available to, e.g., data replication and management middlewares.

%% file: sections/4_scenario.tex
\section{Scenario}
\label{sec:scenario}

We adapt our benchmark scenario from the volcano monitoring sensor network described by Werner-Allen et al.~\cite{werner2006fidelity}.
Organizations such as the \ac{usgs}~\cite{usgs} monitor activity of active and dormant volcanoes.
The \ac{hvo}~\cite{hvo} focuses on volcanoes on the Hawaiian Islands, recording and analyzing geological activity, assessing risks, and issuing warnings.

\begin{figure}
    \centering
    \includegraphics[width=\linewidth]{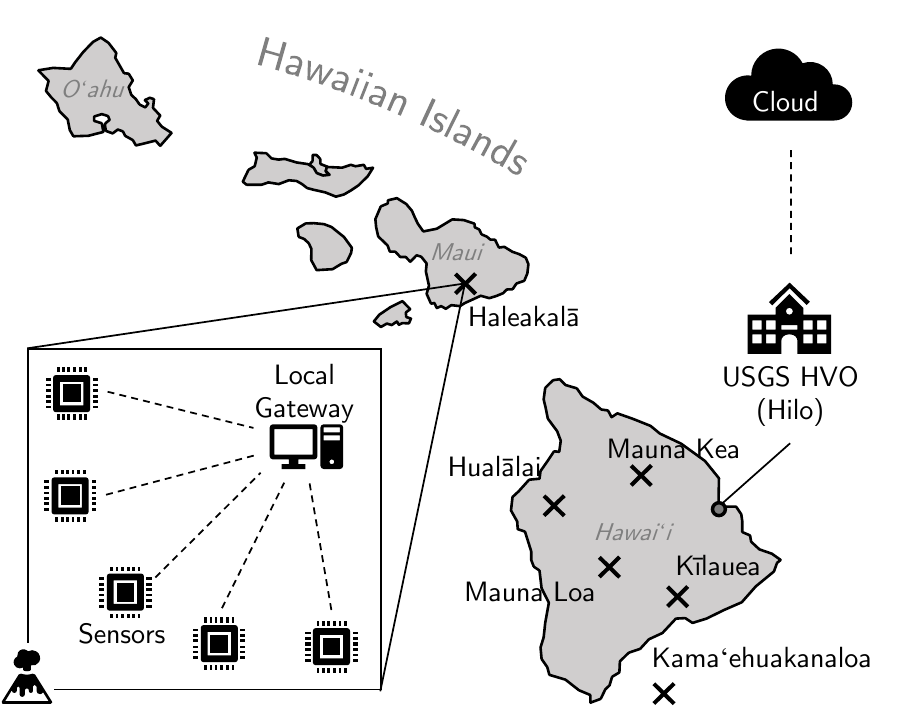}
    \caption{The benchmark scenario is based on an \ac{IoT} sensor network used to monitor six volcanic sites on the Hawaiian Islands. Each site is equipped with multiple sensors, connected to a local gateway. Gateways send data to the \ac{hvo} office in Hilo, Hawaii. A cloud backend is used to store historical data.}
    \label{fig:map}
\end{figure}

It is feasible that the \ac{hvo} would adopt a sensor network as shown in \cref{fig:map} to record activity of the six active volcanos in the region~\cite{10.1145/3487942}.
Similar to the sensor network presented in~\cite{werner2006fidelity}, several sensors with local processing capabilities are located at each site, connected to a local gateway over a radio access network, e.g., LoRaWAN~\cite{lorawan,8019271}.
These sensors produce data at such high rates that complete transmission could exceed LoRaWAN bandwidth or energy availability, hence data will often be pre-processed for anomaly detection locally and only anomalous data is transferred to the local gateway.
In line with the concepts of fog computing, this local gateway is also equipped with processing capabilities.
Further, the local gateways are connected to the \ac{hvo} office in Hilo, Hawaii, where all sensor data may be aggregated and analyzed locally in a data center.
This connection can be over fiber for volcanic sites close to inhabited areas, such as near K\=\i{}lauea, or a radio-based connection such as LTE-M~\cite{lauridsen2016coverage} for remote sites such as Kamaʻehuakanaloa, which is 30km from shore.
Finally, a cloud backend, e.g., provided by the \ac{usgs}, is used to store historic data and serve reports.

\begin{figure}
    \centering
    \includegraphics[width=\linewidth]{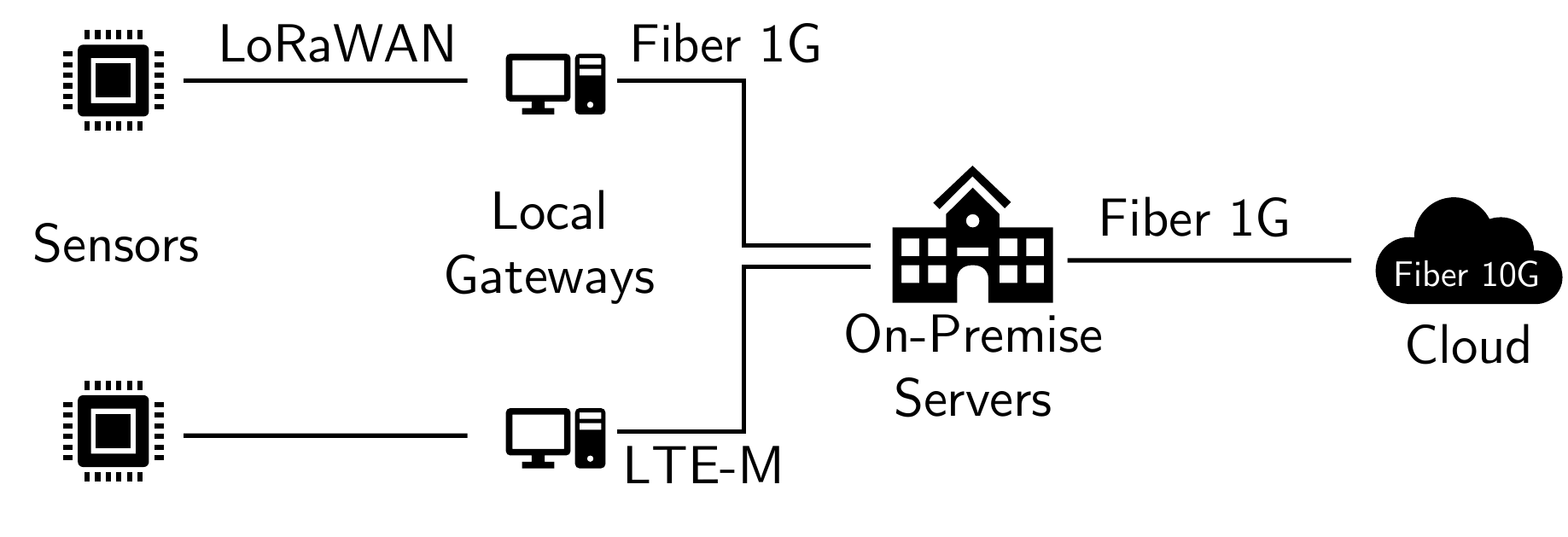}
    \caption{The infrastructure used in our scenario is typical for fog computing, with a large amount of edge devices and multiple layers of compute spread across a large geographical area.}
    \label{fig:overview}
\end{figure}

We show the resulting fog infrastructure in \cref{fig:overview}.
While the scenario may be hypothetical, it combines all challenges and paradigms of data processing in fog computing:
(i)~There is a high volume of data ingress, with a large amount of sensors at the edge, (ii)~local edge processing must be used to save radio bandwidth and device energy, (iii)~there is a need for event loops at the edge, to react to events with low latency, (iv)~data must be managed across a large geographical area, (v)~live transactional workloads on the data are performed within the fog network, (vi)~long-term data storage and analytical workloads are performed in the cloud~\cite{paper_pfandzelter_zero2fog,paper_pfandzelter_functions_vs_streams}.

Interestingly, this scenario also shows how different services of a fog application also have differing dimensions of quality in regard to a fog data middleware.
At the edge, with low-latency event processing loops, the goal is to achieve a processing \ac{SLO} using as few resources as possible, reducing energy consumption and cost~\cite{10.1145/3447545.3451190,HENNING2021100209,cao2022microedge}.
For the data analysis services, however, we see common cloud goals of reducing processing latency and serving data at scale.

%% file: sections/5_benchmark.tex
\section{Benchmark Design}
\label{sec:benchmark}

\begin{figure}
    \centering
    \includegraphics[width=\linewidth]{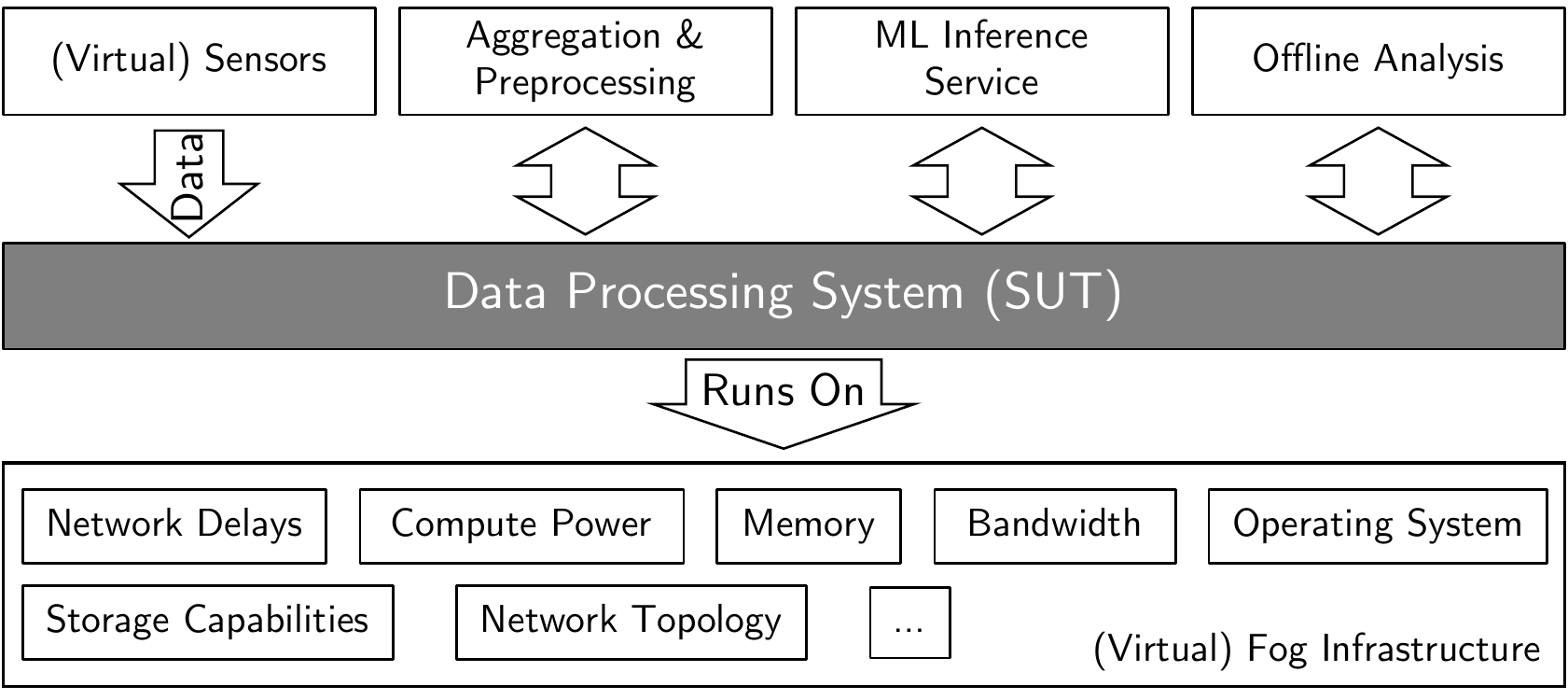}
    \caption{The benchmark sits around the central \ac{SUT}, which is the fog data processing system that is benchmarked. On the one hand, the workload application services run against and partly within the data processing system. All software is deployed onto a (virtual) fog infrastructure, the specification of which is part of the benchmark.}
    \label{fig:architecture}
\end{figure}

We show the overall architecture of the benchmark in \cref{fig:architecture}.
Around the central \ac{SUT}, the data processing system, there are three main components of the benchmark specification that we describe in more detail in this section:
The workload specification (\cref{sec:benchmark:workload}), the infrastructure specification (\cref{sec:benchmark:infrastructure}), and the benchmark metrics that quantify the quality dimensions of the \ac{SUT} (\cref{sec:benchmark:metrics}).

\begin{table}
    \renewcommand{\arraystretch}{1.2}
    \caption{Workload Parameters (Variable Parameters in Bold)}
    \label{tab:workloadparameters}
    \centering
    \begin{tabular}{lp{0.5\linewidth}r}
        \toprule
        Parameter                            & Description                                                                    & (Default) Value        \\
        \midrule
        $N_\mathrm{s}$                       & number of sensors per volcano                                                  & $300$                  \\
        $S_{b}$                              & size of local sensor buffer locally                                            & $1000$                 \\
        ${N}_{\mathit{agg}}$                 & number of samples combined in aggregate reports                                & $25$                   \\
        $y$                                  & ratio of sensors that report an event for gateway to forward event information & $20\%$                 \\
        $r$                                  & sensor sample resolution                                                       & $3\times 64\text{bit}$ \\
        $f$                                  & sensor sampling rate                                                           & $100\text{Hz}$         \\
        $p$                                  & probability a sensor reading exceeds its threshold                             & $15\%$                 \\
        $t_\mathrm{LSTM}$                    & LSTM input data size                                                           & $10\text{s}$           \\
        $q_\mathrm{recent}$                  & ratio of most recent 1h interval reads                                         & $50\%$                 \\
        $q_\mathrm{random}$                  & ratio of random 1h interval reads                                              & $30\%$                 \\
        $q_\mathrm{scan}$                    & ratio of scan/filter read requests                                             & $20\%$                 \\
        $N_\text{Clients}$                   & offline parallel clients
                                             & 100                                                                                                     \\
        \textbf{$\textit{R}_\text{Request}$} & \textbf{offline client request rate}                                           & \textbf{1Hz}           \\
        $t_\mathrm{stale}$                   & offline data staleness threshold                                               & $5\text{s}$            \\
        $N_\mathrm{cloud}$                   & number of cloud instances                                                      & $3$                    \\
        \bottomrule
    \end{tabular}
\end{table}

\begin{table}
    \renewcommand{\arraystretch}{1.2}
    \caption{Compute Parameters (Variable Parameters in Bold)}
    \label{tab:compute_parameters}
    \centering
    \begin{tabular}{lrrr}
        \toprule
        Component ($C$)       & $\mathrm{CPU}_C$   & $\mathrm{Mem}_C$ & $\mathrm{Disk}_C$ \\
        \midrule
        \textbf{Sensor ($s$)} & \textbf{0.25~core} & \textbf{256MB}   & \textbf{4GB}      \\
        Local Gateway ($gw$)  & $4~\text{cores}$   & $4\text{GB}$     & $256\text{GB}$    \\
        On-Premise ($onprem$) & $32~\text{cores}$  & $48\text{GB}$    & $2\text{TB}$      \\
        Cloud ($cloud$)       & $48~\text{cores}$  & $96\text{GB}$    & $4\text{TB}$      \\
        \bottomrule
    \end{tabular}
\end{table}

\begin{table*}
    \renewcommand{\arraystretch}{1.2}
    \caption{Network Infrastructure Parameters}
    \label{tab:network_parameters}
    \centering
    \begin{tabular}{crrrrrrr}
        \toprule
        \multicolumn{1}{c}{\begin{tabular}[c]{@{}c@{}}Connection\\Type\end{tabular}} & \multicolumn{1}{c}{\begin{tabular}[c]{@{}c@{}}Propagation\\Delay\end{tabular}} & \multicolumn{1}{c}{\begin{tabular}[c]{@{}c@{}}$\sigma$ Propagation\\Delay\end{tabular}} & Bandwidth         & \multicolumn{1}{c}{\begin{tabular}[c]{@{}c@{}}Packet\\Loss\end{tabular}} & \multicolumn{1}{c}{\begin{tabular}[c]{@{}c@{}}Packet\\Corruption\end{tabular}} & \multicolumn{1}{c}{\begin{tabular}[c]{@{}c@{}}Packet\\Reordering\end{tabular}} & \multicolumn{1}{c}{\begin{tabular}[c]{@{}c@{}}Packet\\Duplication\end{tabular}} \\
        \midrule

        LoRaWAN                                                                      & $0.021\mathrm{ms}/\mathrm{km}$                                                 & $10\%$                                                                                  & $22\mathrm{kbps}$ & $5\%$                                                                    & $5\%$                                                                          & $5\%$                                                                          & $5\%$                                                                           \\
        LTE-M                                                                        & $0.017\mathrm{ms}/\mathrm{km}$                                                 & $10\%$                                                                                  & $1\mathrm{Mbps}$  & $1\%$                                                                    & $1\%$                                                                          & $1\%$                                                                          & $1\%$                                                                           \\
        Fiber 1G                                                                     & $0.0085\mathrm{ms}/\mathrm{km}$                                                & $10\%$                                                                                  & $1\mathrm{Gbps}$  & $0.1\%$                                                                  & $0.1\%$                                                                        & $0.1\%$                                                                        & $0.1\%$                                                                         \\
        Fiber 10G                                                                    & $0.0085\mathrm{ms}/\mathrm{km}$                                                & $10\%$                                                                                  & $1\mathrm{Gbps}$  & $0.1\%$                                                                  & $0.1\%$                                                                        & $0.1\%$                                                                        & $0.1\%$                                                                         \\

        \bottomrule
    \end{tabular}
\end{table*}

\subsection{Workloads}
\label{sec:benchmark:workload}

\begin{figure}
    \centering
    \includegraphics[width=\linewidth]{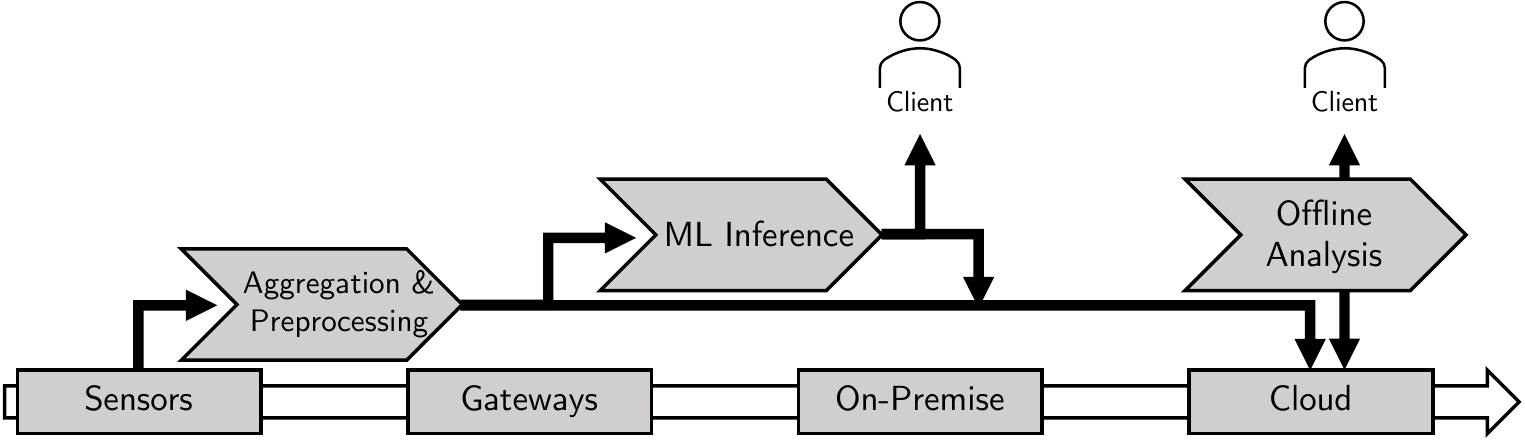}
    \caption{The application logic comprises three services: \emph{Aggregation \& Preprocessing} aggregates sensors data and handles event detection; \emph{\ac{ML} Inference} runs an \ac{LSTM} model detecting anomalous sensor readings; and in \emph{Offline Analysis} clients request historic time-series sensor records.}
    \label{fig:dataflow}
\end{figure}

Our benchmark workload comprises three services at different stages of the edge-cloud continuum (\emph{Objective~1}), as illustrated in \cref{fig:dataflow}.
Note that our workloads merely specify \emph{functionality}, not actual service implementations (\emph{Objective~3}).

The online filtering operations and online monitoring services rely on a data ingress point at the edge.
Here, based on the scenario we presented in \cref{sec:scenario}, sensors produce frequent monitoring data.
Sensors locally store a buffer of $S_{b}$ sensor readings.
Further, two algorithms typical of edge data preprocessing are computed near the edge:
First, every $N_{\mathit{agg}}$ sensor readings, aggregate average sensor readings are generated.
Only these aggregate readings are sent to the following services.
Second, if a sensor reading exceeds a threshold, which happens with probability $p$, an event is immediately sent to the edge service.
If more than $y\%$ of sensors of one location report that their sensor reading threshold has been exceeded, the edge service collects buffered data from all local sensors and reports it upstream.
Since the rate of sensor readings is fixed, this workload follows an open system model~\cite{schroeder2006open}.
In addition to the mentioned parameters, the sampling rate $f$, sensor resolution $r$, and number of sensors $N_\mathrm{s}$ all impact the workload on subsequent processing steps, including the offline analysis.
We give a complete list of (variable) workload parameters in \cref{tab:workloadparameters}.

Note that some system properties are derived directly from these parameters.
For example, the theoretical (without data structure overhead) rate of data generated by a sensor is $r \times f$.
With the parameter values given in \cref{tab:workloadparameters}, a sensor generates 19.2kbit/s of data.
The mean bandwidth used between a sensor and the edge service is:

\begin{equation*}
    \overline{d_{\mathrm{s \to gw}}} = \frac{r \times f}{N_\mathrm{agg}} + (1 - \sum_{k=0}^{\lceil N \times y \rceil} {N\choose k}p^k(1-p)^{N-k}) \times r \times S_{b} \times f
\end{equation*}

, i.e., the rate of regular aggregated sensor readings and the rate of infrequent events by multiple sensors at a location.
With the given values, the mean bandwidth between a sensor and the edge service is 149.5kbit/s.
Consequently, each edge service processes a mean 44.84Mbit/s of data.

Both regularly aggregated and event-based data is forwarded to the intermediate ML inference service.
The live data processing here comprises augmenting data from different source volcano locations with inference results from a \ac{ML} model.
Specifically, aggregated data records should be used for time-series forecasting using a \ac{LSTM} model to annotate each record with a probability of a major seismic event at this location, taking into account the most recent $t_\mathrm{LSTM}$ of data for that location~\cite{mousavi2020earthquake,10.1145/3386723.3387865} and generating warnings in Hilo.
Annotated results are then forwarded to the offline analysis service.

The offline analysis workload comprises common time-series database operations.
Insert operations are performed by means of incoming data from the fog side, i.e., aggregated sensor records with \ac{ML} inference annotations.
Read operations are performed by clients in three distinct operations: batch reads of the most recent one-hour interval ($q_\mathrm{recent}$, $50\%$), batch reads of random one-hour intervals ($q_\mathrm{random}$, $30\%$), and scans with filter operations ($q_\mathrm{scan}$, $20\%$).
This workload follows an open model, i.e., $N_\mathrm{Clients}$ clients send requests at a constant rate of $R_\mathrm{Request}$ rather than waiting for responses, with a high number of concurrent queries by multiple clients.
Geographically, these clients can be considered as located at the cloud location.
In order to precisely measure system performance, we aim for a high (e.g., 80\%) utilization of system resources~\cite{book_cloud_service_benchmarking} and the $R_\mathrm{Request}$ parameter should be adapted to yield such utilization.
In order to derive meaningful and representative metrics for this offline system, we also introduce a \emph{staleness threshold} ($t_\mathrm{stale}$) for data, which is more meaningful for an offline system than, e.g., a precise measurement of end-to-end latency (\cref{sec:benchmark:metrics}).

\subsection{Infrastructure}
\label{sec:benchmark:infrastructure}

As outlined in \emph{Objective~2}, functionality and performance of a fog application not only depend on the workload used but also on the underlying infrastructure and network topology.
We thus also provide an infrastructure specification in our benchmark, based on the topology shown in \cref{fig:overview}.
The complete list of infrastructure parameters is shown in \cref{tab:compute_parameters,tab:network_parameters} for compute and network infrastructure, respectively.

For every type of component $C$, e.g., the sensor board, CPU type ($\mathrm{CPU}_{C}$), memory size ($\mathrm{Mem}_{C}$), disk size ($\mathrm{Disk}_{C}$), and operating system may be changed.
Further, in the cloud backend, multiple ($N_\mathrm{cloud}$) identical machines are combined into a cluster.
The network parameters include a propagation delay, standard deviation of propagation, delay, link bandwidth, packet loss rate, packet corruption rate, packet reordering rate, and packet duplication rate for different link types based on MockFog (see the implementation description in \cref{sec:implementation}).

\subsection{Metrics}
\label{sec:benchmark:metrics}

As we combine workloads from different paradigms, the metrics we use to evaluate an \ac{SUT} reflect the different desirable dimensions in fog data processing.
We hence propose distinct metrics for each of the three workloads.

The edge sensor data aggregation and preprocessing workload is \ac{SLO}-driven, i.e., data must be processed at the rate with which it is generated in order to prevent queuing.
A meaningful metric is thus the amount of edge compute resources (and, by extension, theoretical cost and energy consumption) with which an \ac{SUT} can achieve this \ac{SLO}~\cite{10.1145/3447545.3451190,HENNING2021100209,cao2022microedge}.
This requires varying the $\mathrm{CPU}_\mathrm{s}$, $\mathrm{Mem}_\mathrm{s}$, and $\mathrm{Disk}_\mathrm{s}$ parameters while observing that processing queuing remains constant.

For online intermediate data processing, the relevant metric is the added latency of shipping data to and from the \ac{ML} inference service and processing it.
An adequate metric is the end-to-end latency between creating an (aggregated) data point on the sensor and its insertion into the cloud offline data store.
To support a more accurate relative comparison of results of different \ac{SUT}, we propose to subtract the network delay imposed by the infrastructure between sensors and cloud, which is a constant offset.

As discussed, edge-to-cloud latency is a misleading metric for offline data analysis as differences in latency are often not noticeable for clients~\cite{mohan2020pruning}.
A more meaningful perspective is that of data staleness violations:
Of all read requests that semantically ask for the \emph{most recent} data, what ratio does not meet the staleness threshold $t_\mathrm{stale}$, i.e., how often is data processing and replication too slow?
The second cloud metric is the latency of requests sent by cloud-based clients.

\section{Implementation}
\label{sec:implementation}

While most of the benchmark implementation is highly specific to an \ac{SUT}, e.g., certain platforms have limitations on programming languages and deployment models that can be used to implement and deploy a service, we provide a number of artifacts to support this implementation.
We make these artifacts available as open-source software.\footnote{\url{https://github.com/OpenFogStack/fog-data-benchmark-specification}}
Our infrastructure specification is implemented in terms of a MockFog~\cite{paper_hasenburg_mockfog,paper_hasenburg_mockfog2} specification, allowing researchers to replicate an emulated infrastructure that meets the specifications outlined in \cref{sec:benchmark}.
Further, we provide a data generator implementation, that accepts the generator parameters outlined in our workload specification, and a re-usable \emph{TensorFlow} implementation of the \ac{LSTM} model used in the ML inference service.

%% file: sections/6_conclusion.tex
\section{Conclusion}
\label{sec:conclusion}

In this paper, we have outlined a benchmark for fog data processing based on an \ac{IoT} sensor network scenario.
Reflecting the reality of fog applications, our benchmark combines geo-distributed workloads of multiple paradigms.
Further, we couple workload and infrastructure specification in our benchmark, as fog application performance depends on fog infrastructure characteristics.
Our benchmark specification is portable across organizational paradigms, allowing researchers to quantitatively compare different approaches to managing and building applications in the fog.
In future work, we plan to use our benchmark to compare such existing approaches from fog computing research, analyze their limits and constraints, and derive gaps in existing research.